\documentclass[12pt]{article}
\usepackage{amsmath,amssymb}

\newcommand{\Ut}[1]{\raisebox{0.ex}[-.2ex][0.ex]{$\overset{\thicksim}{\rule{0mm}{1.9mm}\smash[t]{#1}\,}$}\mbox{}\!}

\newcommand{\ut}[1]{\overset{\thicksim}{\rule{0mm}{1.1mm}\smash[t]{#1}}\mbox{}}

\def\beq#1\eneq{\begin{equation}#1\end{equation}}
\newcommand{\bqa}{\begin{eqnarray}}
\newcommand{\enqa}{\end{eqnarray}}
\newcommand{\bqann}{\begin{eqnarray*}}
\newcommand{\enqann}{\end{eqnarray*}}

\begin{document}
\centerline{\Large {\bf 2-Form Gravity of the Lorentzian Signature}}
\vskip 1\baselineskip

\centerline{Jerzy Lewandowski\footnote{lewand@fuw.edu.pl} and Andrzej 
Oko{\l}{\'o}w\footnote{oko@fuw.edu.pl}}  
\medskip

\centerline{\it Instytut Fizyki Teoretycznej, Uniwersytet Warszawski,} 
\centerline{\it ul. Ho\.za 69, 00-681 Warszawa, Poland}                
\bigskip

\abstract{We introduce a new spinorial, BF-like action for 
the Einstein gravity. This is a first, up to our knowledge,
2-form action which describes the real, Lorentzian gravity and 
uses only the self-dual connection. In the generic case, the 
corresponding classical canonical theory is equivalent to the  
Einstein-Ashtekar theory plus the reality conditions.} 

\newpage

\noindent {\bf Introduction.} It is well known that, in four dimensions, 
a metric tensor can 
be derived from a suitably normalized triad of self-dual 2-forms
\cite{ProfessorPlebanski,Urbantke,CDJ,CDJM,Ingemar,Yura,Smolin}.
Given a complex 4 dimensional vector space $V$  equipped
with a complex valued metric tensor,  one decomposes the
space $\wedge^2 V^{\star}$ of two forms into the direct,
wedge-orthogonal sum 
\[
\wedge^{2} V^{\star} = \Omega^{-} \oplus \Omega^{+}
\]  
of the spaces of anti-self-dual and self-dual 2-forms, respectively. 
The space $\Omega^+$ (as well as $\Omega^-$), determines the metric 
tensor up to a conformal factor. The missing information about 
the metric tensor  can be provided by indicating an 
element ${\Sigma^0}_{0} \in  \Omega^+$ such that
\[
{\Sigma^0}_{0}\wedge {\Sigma^0}_{0} = i\,{\rm vol},
\]
where `vol' stands for the volume 4-form defined by the
metric tensor.
This property gave motivation to formulate Einstein's gravity
as a theory of the 2-forms rather then the metric tensors.   
In the cases of the complex gravity, and the real gravity of the 
Euclidean signature, there is known an action, introduced by 
Capovilla, Dell and Jacobson (CDJ),
which is written purely in terms of the self-dual connection 
and self-dual 2-forms. In the real Lorentzian case, however,
the actions which are known, are either Pleba{\'n}ski's action
which involves both self-dual and anti-self-dual
connections and 2-forms or one uses the complex version of
the CDJ action plus extra reality conditions imposed on the 
solutions. The canonical theory based on the CDJ action is 
equivalent to the Ashtekar \cite{Abhay1,Abhay2} theory,
where again the reality conditions are  taken into account
as some extra conditions. Recently, the 2-form gravity is often 
viewed as a close neighbour of the so called BF theory and is 
applied in the spin-foam quantization \cite{Carlo,Baez}. 
There are also attempts to use the 2-form approach
of general relativity in a construction of a holographic 
formulation of gravity \cite{Smolin}. 

In this letter, we focus on the Einstein gravity of the 
Lorentzian signature. We introduce an action which involves 
only two forms and self-dual connections and incorporates all 
the reality conditions. To our knowledge this is the first 
action of those properties, although it seems  quite geometric
and natural.  The canonical theory derived from this action \cite{Okolow1}  
will be described in detail in a subsequent paper \cite{Okolow2}. 
Here, we will briefly outline the results.
\bigskip

\noindent{\bf The action.}
Let $M$ be a four dimensional real manifold. The main variables of 
our theory are a 1-form $A$ and a 2-form $\Sigma$ both defined on $M$
and taking values in the Lie algebra  $sl(2,\mathbb{C})$ of the group  
$SL(2,\mathbb{C})$.
The group acts in a certain abstract spinor space $S$ and preserves
the antisymmetric (symplectic) 2-form $\epsilon$ defined therein.
The complex conjugation of spinors is an anti-linear isomorphisms that carries $S$ into another spinor space $S'$ 
\[
S\ni \mu\ \mapsto\ \mu^{\ast} \in S', 
\]
and the symplectic form
$\epsilon$ into the symplectic form $\epsilon'$ of $S'$.

The action is defined as follows\footnote{We use here the standard notation \cite{Roger}, expect indication of complex conjugation of spinors --- we mark it by means of $\ast$. 
The spinors  are represented by their components, with respect to basis 
$o, \, \iota\in S$, and $o^{\ast},\, \iota^{\ast}\in S'$, that is
$\mu\ =\ \mu^0 o\  +\ \mu^1\iota \in S$, and   
$\nu'\ =\ \nu^{0'} o^{\ast}  +\ \nu^{1'}\iota^{\ast}\in S'$.
The indices are lowered and raised by the symplectic forms, that is
$\mu_{A}=\mu^{B}\epsilon_{BA},\ \mu^{A}=\mu_{B}\epsilon^{AB},\ 
\nu_{A'}=\nu^{B'}{\epsilon'}_{B'A'},\ \nu^{A'}=\nu_{B'}{\epsilon'}^{A'B'}$
and finally $\epsilon^{AC}\epsilon_{BC}=\delta^{A}_{B},\ 
{\epsilon'}^{A'C'}\epsilon_{B'C'}=\delta^{A'}_{B'}$.}:
\begin{multline}
S[\Sigma^{AB},\,A_{AB},\,\Psi_{ABCD},\,\Phi_{ABC'D'},\, R]=
\\ =\int \Sigma^{AB}\wedge F_{AB} -
\frac{1}{2}\Psi_{ABCD}\Sigma^{AB} \wedge\Sigma^{CD}-\label{dzialanie}\\
 - \Phi_{ABC'D'} \Sigma^{AB} \wedge \Sigma^{\ast C'D'} -
\frac{1}{2}R ( \Sigma^{AB} \wedge \Sigma_{AB} + \Sigma^{\ast A'B'}
\wedge \Sigma^{\ast}_{ A'B'})
\end{multline}
where
\[
F_{AB} := dA_{AB} +A_{A}\!^{C}\wedge A_{CB},
\]
and the extra spinor fields $\Psi$ and $\Phi$, and the scalar field $R$  play 
the role of the Lagrange multipliers which satisfy\footnote{Since 
$A$ and $\Sigma$ take values in  $sl(2,\mathbb{C})$, $A_{AB} = A_{(AB)}$
and $\Sigma^{AB}=\Sigma^{(AB)}$.}
\[
\Psi_{ABCD}\ =\ \Psi_{(ABCD)},\ \Phi_{ABC'D'}\ =\ \Phi_{(AB)(C'D')}.
\]

The vanishing of the variations of the action $S$ with respect to 
$\Psi
,\, \Phi,\, R$ is, generically (we indicate the limitations
below),  equivalent to the existence and uniqueness
of the corresponding metric tensor, real and of the Lorentzian
signature $\pm(+---)$. Indeed:  
\begin{eqnarray}
\partial_{\Psi} S=0 & \Rightarrow &  \Sigma^{(AB}\wedge \Sigma^{CD)}= 0;
\label{c1}\\
\partial_{\Phi} S = 0 & \Rightarrow& \Sigma^{AB}\wedge 
\Sigma^{\ast C'D'} = 0;\label{c2}\\
\partial_{R} S =0 &\Rightarrow & \Sigma^{AB}\wedge \Sigma_{AB} + 
\Sigma^{\ast A'B'}\wedge \Sigma^{\ast}_{A'B'} = 0.\label{c3}
\end{eqnarray}
The first condition, generically,  implies  the existence of four
linearly independent complex valued 1-forms 
$\theta^{AB'},\,A,B=0,1$ tangent to $M$,  such that    
\[
\Sigma^{AB}=\theta^{AA'}\wedge\theta^{BB'}\epsilon_{B'A'}.
\]
Due to the conditions (\ref{c2}) and (\ref{c3}) the metric tensor
defined by the above null frame, that is
\[
g\ :=\ \theta^{AC'}\otimes \theta^{BD'}\epsilon_{AB}\epsilon_{C'D'}
\]
is real and its signature is  $(+---)$ or $(-+++)$. 
 The above conditions, and the reconstruction 
of the metric tensor come from Pleba{\'n}ski \cite{ProfessorPlebanski} (see also
\cite{CDJM}).

Given the metric tensor, the vanishing of the variation of $S$ with  
respect to $A$ implies the metricity of $A$, 
\begin{equation} \label{metr}  
\partial_A S\ =\ 0\ \Rightarrow\ d\Sigma^{AB}\ - A^{A}\!_{C}\wedge\Sigma^{CB}\
- A^{B}\!_{C}\wedge\Sigma^{AC} = 0. 
\end{equation}
From this equation, the 1-forms ${A^A}_B$ are determined completely 
by $\Sigma$ and set the spin connection corresponding to the spin 
structure defined locally by 
\[
S\otimes S'\ni \mu\otimes \nu'\ \mapsto\ \mu_A\nu_{B'}\theta^{AB'}.    
\]         
At this point, the meaning of the spinors $\Psi$ and $\Phi$ and the scalar 
$R$ becomes clear  
after taking the variation of $S$ with respect to 
$\Sigma$, namely\footnote{This is a holomorphic
variation, that is $\partial_{\Sigma} \Sigma^{\ast}=0$.}:
\[
\partial_\Sigma S \ =\ 0\ \Rightarrow\ 
F_{AB}=\Psi_{ABCD}\Sigma^{CD}+\Phi_{ABC'D'}\Sigma^{\ast C'D'}+
R\Sigma_{AB}.
\]      
This is exactly the familiar spin decomposition of the spinorial
curvature, where $\Psi$ is the Weyl spinor, $\Phi$ represents
the traceless part of the Ricci tensor and $12 R$ is the Ricci scalar.   

The vacuum Einstein equations (imposed on the derived metric) 
follow from the vanishing of the variation with respect to 
$\Sigma^{\ast}$, namely
\begin{equation} \label{E}
\partial_{\Sigma^{\ast}} S \ =\ 0\ \Rightarrow\ \Phi_{ABC'D'}\Sigma^{AB}\ +\ 
R\Sigma^{\ast}_{C'D'}\ =\ 0. 
\end{equation}
Since in the generic case, the six 2-forms $\Sigma^{(AB)},\,
\Sigma^{\ast}_{(C'D')}$ form a basis of the complexified space 
of 2-forms, the above equation implies
\[
\Phi_{ABC'D'}\ =\ 0\ =\ R.
\] 

As emphasized, the above reconstruction of Einstein's
theory applies to the generic case. The genericity condition
is
\beq
\Sigma^{00}\wedge \Sigma^{11}\ \not=\ 0.
\label{gen-cond}
\eneq
Otherwise, for every values of $A,B,C,D$ 
\[
\Sigma^{AB}\wedge\Sigma^{CD}\ =\ 0.
\]
Every degenerate triad of 2-forms $\Sigma^{AB}$ is generated by 
the wedge product and the linear combinations
from certain triad of 1-forms. Farther degeneracy 
takes place if the triad of 1-forms is not linearly independent,
the most degenerate case being just 
\[
\Sigma \ =\ 0.
\]
Given degenerate $\Sigma$, the equations (\ref{metr},\,\ref{E})
do not determine $A,\, \Phi$ and $R$. The freedom depends on the 
degeneracy. For example, in the most degenerate case, the general
solution is: a flat connection $A$ and  arbitrary $\Phi$, $\Psi$ and 
$R$.
\bigskip

\noindent{\bf Hamiltonian and the constraints.}
For the construction of the canonical theory corresponding to the action 
(\ref{dzialanie}) let us introduce a variable $t$
and a coordinates system $(x^0 =t, x^i).$  
Then, the configuration  variables $q^r$ and the corresponding canonically 
conjugate momenta $\ut{p}_r$ are  
\begin{gather*}
\{q^{r}\} = \{  A_{\alpha AB},\Sigma^{AB}_{0i},\Sigma^{AB}_{ij},
\Sigma^{\ast A'B'}_{0i},\Sigma^{\ast A'B'}_{ij},
\Psi_{ABCD},\Phi_{ABC'D'},R\},\\
\{\ut{p}_{r}\}\ =\ \{ \ut{p}^{\alpha AB},\ut{p}_{AB}^{0i},\ut{p}_{AB}^{ij},
\ut{p}_{A'B'}^{\ast 0i},\ut{p}_{A'B'}^{\ast ij},
\ut{p}^{ABCD},\ut{p}^{ABC'D'},\ut{p}_{R}\},
\end{gather*}
where we take into account only independent components 
($(A,B)=(0,0),\, (0,1),\, (1,1)$ in $A_{\alpha AB}$, etc.).
Obviously, the Legendre transform $(q^r,\, \partial_{t}q^{r}) \mapsto (q^r,\,\ut{p}_r)$ 
is  not invertible and gives the primary constraints: 
\beq
\begin{cases}
\Ut{\phi}^{iAB}:=\ut{p}^{iAB}-\ut{\sigma}^{iAB}=0&\\
\Ut{\phi}_{r}:=\ut{p}_{r}=0&\text{otherwise,}
\end{cases}
\label{wiezy}
\eneq
where\footnote{$\ut{\epsilon}^{\alpha\beta\gamma\delta}$ is the Levi-Civita 
density of weight 1 on $\cal M$.}
\[
\ut{\sigma}^{iAB}:=\ut{\epsilon}^{ijk}\Sigma^{AB}_{jk};\;\;\;\;\; \ut{\epsilon}^{ijk}:=\ut{\epsilon}^{0ijk}
\]
Therefore, the Hamiltonian is defined up to  an additive term, a combination
of the constraints and Lagrange multipliers, that is  
\begin{multline}
H(q,\ut{p},u)=\int_{S}{\rm Tr}[ \,\ut{\sigma}^{i} D_{i}  A_{0} -
2\Sigma_{0i}\ut{\epsilon}^{ijk}F_{jk}+2\Sigma_{0i}\Psi \, \ut{\sigma}^{i}+
\\+ 2\Sigma_{0i}\Phi \,
\ut{\sigma}^{\ast i}+2 \, \ut{\sigma}^{i}\Phi\Sigma^{\ast}_{0i}+
2R(\Sigma_{0i} \, \ut{\sigma}^{i} + \Sigma^{\ast}_{0i} \,
\ut{\sigma}^{\ast i}) \, ] +\int_{S}u^{r}\Ut{\phi}_{r}
\label{ham'-1}
\end{multline} 
where:
\[
D_{i}A_{0AB}:=\partial_{i}A_{AB}+A_{iA}\!^{C}A_{0CB}+A_{iB}\!^{C}A_{0AC}
\]
and the Lagrange multipliers $u^{r}$ are tensors on $S$ which have the same 
(anti)symmetries in the indices as the corresponding primary constraint functions 
$\Ut{\phi}_{r}$. 

Demanding that  the primary constraints $\Ut{\phi}$ are preserved during the time 
evolution, that is: 
\beq
\{...\{\{\Ut{\phi},H\},H\}, ..., H\}\ =\ 0,
\label{time-der-of-constr}
\eneq
one gets  secondary constraints and  restrictions on the Lagrange multipliers. 
The complete set of the constraints for the Hamiltonian (\ref{ham'-1})
is known in the case, when the triad $(\ut{\sigma}^{i00},\,
\ut{\sigma}^{i01},\,\ut{\sigma}^{i11})$ of the vector densities
is linearly independent, that is,  when the 2-forms $\Sigma^{AB}$
give rise to a Lorentzian metric tensor  in the space-time 
$M$  and the three-metric induced on the  $t={\rm const}$ surfaces
 is not degenerate (the signature of the induced three-metric is $\pm(+++)$ or $\pm(-++)$). Then the set of the constraints
consists of the primary ones (\ref{wiezy}) and of the following 
secondary constraints:
\begin{gather}
\Sigma^{(AB}_{0i}\ut{\sigma}^{iCD)}=0;\;\;
\Phi_{ABC'D'}=0;\;\;
R=0 \nonumber\\
D_{i}\ut{\sigma}^{iAB}=0;\;\;
\epsilon^{ijk}F_{jkAB}-\Psi_{ABCD}\ut{\sigma}^{iCD}=0\nonumber\\
\Sigma^{AB}_{0i}\ut{\sigma}^{\ast iC'D'}+
\Sigma^{\ast C'D'}_{0i}\ut{\sigma}^{iAB}=0\\
\Sigma^{AB}_{0i}\ut{\sigma}^{i}_{AB}+\Sigma^{\ast A'B'}_{0i}
\ut{\sigma}^{\ast i}_{A'B'}=0\nonumber\\
D_{k}(\ut{\sigma}^{kCA}\ut{\sigma}^{(iB}\!_{C})\ut{\sigma}^{j)}\!_{AB}+
[D_{k}(\ut{\sigma}^{kCA}\ut{\sigma}^{(iB}\!_{C})
\ut{\sigma}^{j)}\!_{AB}]^{\ast}=0 \nonumber
\end{gather} 

Finally, the conditions on the Lagrange multipliers $u$
implied by (\ref{time-der-of-constr}) are:
\begin{gather*}
A_{0C}\!^{(A}\ut{\sigma}^{iB)C}+D_{j}(\ut{\epsilon}^{ijk}\Sigma_{0k}^{AB})-\ut{\epsilon}^{ijk}u^{AB}_{jk}=0\\
D_{k} A_{0AB}+
2\Psi_{ABCD}\Sigma^{CD}_{0k}-u_{kAB}=0\\
\ut{\epsilon}^{ijk}(D_{j}u_{kAB}-2\Psi_{ABCD}u_{jk}^{CD})-u_{ABCD}\ut{\sigma}^{iCD}=0\\
u_{R}=0;\;\; u_{ABC'D'}=0;\;\;\; u^{(AB}_{0i}\ut{\sigma}^{iCD)}+
2\Sigma^{(AB}_{0i}\ut{\epsilon}^{ijk}u_{jk}^{CD)}=0\\
u^{AB}_{0i}\ut{\sigma}^{\ast iC'D'}+
2\Sigma^{AB}_{0i}\ut{\epsilon}^{ijk}u_{jk}^{\ast C'D'}+u^{\ast C'D'}_{0i}\ut{\sigma}^{iAB}+
2\Sigma^{\ast C'D'}_{0i}\ut{\epsilon}^{ijk}u_{jk}^{AB}=0\\
u^{AB}_{0i}\ut{\sigma}^{i}\!_{AB}+2\Sigma^{AB}_{0i}\ut{\epsilon}^{ijk}u_{jkAB}+
u^{\ast A'B'}_{0i}\ut{\sigma}^{\ast i}\!_{A'B'}+
2\Sigma^{\ast A'B'}_{0i}\ut{\epsilon}^{ijk}u^{\ast}_{jkA'B'}=0
\end{gather*}

Concluding,  the classical theory given by our action is 
equivalent to the real section of the Ashtekar theory
as long as  the triads $(\ut{\sigma}^{i00},\,
\ut{\sigma}^{i01},\,\ut{\sigma}^{i11})$ are linearly independent. In the degenerate case,
however, the theories are different \cite{Okolow1,Okolow2}. Our formulation 
provides a new starting point for the quantization.
Then, the differences in the degenerate sector may become
relevant, because the quantum geometry \cite{AL} is degenerate in 
most of the space points.

{\bf Acknowledgements.}
We thank Abhay Ashtekar, John Baez, Ingemar Bengtsson, Jerzy Kijowski
and Lee Smolin for their comments. 
JL thanks MPI for Gravitational Physics in Potsdam-Golm and the 
organizers of the workshop Strong Gravitational 
Fields held in Santa Barbara for their hospitality. This research was 
supported in part by Albert Einstein MPI, University of Santa Barbara,
and  the Polish Committee for Scientific Research under  grant no. 2 
P03B 060 17.

\end{document}